\documentclass[preprint]{elsarticle}
\usepackage{amssymb}

\usepackage{amsmath}
\usepackage{graphics}
\usepackage{epsfig}
\usepackage{lipsum}
\usepackage{comment}
\usepackage{tikz}

\usepackage{tkz-graph, tikz-cd}
\usetikzlibrary{calc,arrows,shapes,snakes,automata,backgrounds,petri}
\usetikzlibrary{decorations.markings}
\usepackage{hyphenat}
\usetikzlibrary{automata}
\usetikzlibrary{positioning}
\usetikzlibrary{topaths}
\usepackage{pstricks}
\usetikzlibrary{automata,positioning,calc}
\usetikzlibrary{arrows}
\usepackage{wasysym} 
\usepackage{color}

\definecolor{verde}{rgb}{0.1,0.5,0.3}
\definecolor{rojo}{rgb}{0.8,0.3,0.3}

\usepackage{relsize}
\usepackage{ctable}
\usepackage{epstopdf}

\usepackage{upgreek}
\usepackage{mathrsfs}

\begin{document}

\title{Markovian simulation for ancestors trees}%


\author[rvt]{C. Jarne}
\ead{cecilia.jarne@unq.edu.ar}
\author[focal]{ M. Caruso}
\ead{mcaruso@ugr.es}

\address[rvt]{ Universidad Nacional de Quilmes - Departamento de Ciencia y Tecnolog\'ia CONICET}
\address[focal]{Departamento de F\'isica Te\'orica y del Cosmos, Universidad de Granada,Campus de Fuentenueva, Spain}


\begin{abstract}
We present a computational model to reconstruct trees of ancestors for animals with sexual reproduction. Through a recursive algorithm combined with a random number generator, it is possible to reproduce the number of ancestors for each generation and use it to constraint the maximum number of the following generation. This new model allows to consider the reproductive preferences of particular species and combine several trees to simulate the behavior of a population. It is also possible to obtain a description analytically, considering the simulation as a theoretical stochastic process. Such process can be generalized in order to use an algorithm associated with it to simulate other similar processes of stochastic nature. The simulation is based in the theoretical model previously presented in \cite{nuestro-paper}.
\end{abstract}


\maketitle

\section{Introduction}

\subsection{On the motivation of the numerical simulation}

There are some previous works in the literature that were originally motivated by the divergence of the geometric model for sexual reproduction of individuals. This means that the number of ancestors in those models is given by a power of 2. In few generations this number reaches arbitrary large values. Those works have shown that it is not possible to reconstruct the genealogical tree of ancestors using a geometric progression \cite{maltusian}. A simple deterministic rule could not reproduce what happens with real animal reproduction where other factors intervene.

In other more realistic models that consider additional factors, the effort focuses on the estimation for distributions of ancestor's repetition along genealogical trees \cite{Derrida-1,Derrida-2,Derrida-3,Douglas, Kelleher} and not over the number of ancestors or the distribution of them as in our previous theoretical work \cite{nuestro-paper}. The last two topics are the objects of our study, but now considering a different perspective.

It is well known that numerical simulations are tools that have proven to be useful for studying different situations. With a simple algorithm, where different parameters are clearly indicated, it is possible to simulate different scenarios by adjusting the values of the parameters in the model.

This new approach is based on our previous work, where we focus in on how to model the problem theoretically using markovian models. Regarding the model, we postulated that blood relationship between ancestors (inbreeding) is the key to understand the deviation from the deterministic progression given by the geometrical progression. When considering a random variable that represents the number of ancestors who are present in a given generation, the size of the state space depends on each generation. This makes difficult to find an analytical solution. We have shown in \cite{nuestro-paper} a method that partially solves this including two parameters to the problem.

Before starting the description of the new approach, there are two suppositions regarding the biology of the system that we considered. These were used in our previous work and also were used as starting point on the present one. Initially,species described here do not have an specific behavior of sexual partner selection i.e. random mating reproduction \cite{5,6}. This means that in our model the partner could be a kin or not. This case consists of the simplest scenario to perform the calculation. The second assumption is about the size of the population. The distribution of ancestors for a given generation is contained in a population large enough to avoid forcing the selection of a kin sexual partners. Mates could be blood related or not at random, up to a maximum generation when the process start to decrease. The inbreeding is caused by the animal behavior (random or directed). In this way, this assumptions are common to develop population genetic models \cite{7,8}.	

Here we present a model, which is studied numerically and analytically. We used the model to generate a simulation and obtain the number of ancestors of an individual at successive previous generations considering a sexually reproducing population. The new model allow us to include a set of parameters related to the match preferences of certain groups of animals, also it includes an algorithmic approach that allow us to perform the probabilistic numerical simulation. We used a simple computational model to track back the chain of ancestors and we analyze some example cases. 

This model, is partially open. This means that it is possible to change the pdf according the random numbers are distributed. In our work we chose a particular realization to be able to show some possible results.

We motivate the use of a variable called $r_n$ regarding ancestry number at each generation to model inbreeding in biological population. We are interested in modeling the inbreeding that naturally arises from the reproductive behavior of the specie and not particularly related to the constraint of the maximum population size.

With the computational calculation we can weigh the inbreeding and be able to perform an estimation on the mean number of ancestors by using actual behavioral information of a given specie.  

In the following sections we show how to construct and keep a general algorithm considering inbreeding preferences and also allowing the further reader be able to include his or her own hypothesis or data about animal preferences via the parameters included in the computational model. Along the work we study the kind of results obtained considering different selection of the parameters.	
	
\subsection{Regarding the biological importance of inbreeding}

It is well establish that close inbreeding within species its a behavior that can result in inbreeding depression. This is caused by an increase in homozygosity of recessive, deleterious alleles and the loss of heterosis \cite{4-cichlid-fish}. On the other hand, a behavior that involves an extreme intraspecific outbreeding can also be disadvantageous for some species. This is called outbreeding depression \cite{17-cichlid-fish}. The beneficial gene complexes or local genetic adaptations could be disrupted in this case, caused by individuals with different adaptations for a different environment \cite{18-cichlid-fish}. 

There are studies that support the fact that animals avoid close kin as mating partners \cite{5-cichlid-fish, 19-cichlid-fish, 20-cichlid-fish}. The theory of optimal outbreeding is supported by experimental behavioral studies  \cite{7-cichlid-fish}. Those studies have shown mating preferences for intermediately related individuals \cite{6-cichlid-fish}. Evidence in this direction is based on genetic studies reporting stabilizing selection on genomic divergence in wild populations of animals \cite{21-cichlid-fish} and plants \cite{22-cichlid-fish}.

Empirical studies were conducted regarding inbreeding strategy. Those studies have reported inbreeding tolerance in the wild for different species such as New Zealand robins, \textit{Petroica australis} \cite{11-cell}; bighorn sheep (\textit{Ovis canadensis}) \cite{16-cell}; great tits (\textit{Parus major}) \cite{17-cell} and even inbreeding preference in cichlid fish (\textit{Pelvicachromis taeniatus}) \cite{18-cell}. Besides, recent studies have found evidence of regular incest behavior in wild mammals, even in social species where relatives are spatio-temporally clustered showing that opportunities for inbreeding frequently arise \cite{BIOL LETT}.

There is an interesting concept named Minimum viable population (MVP). It is defined as a lower bound on the population of a species such that it can survive in the wild species. This is a population context-specific concept and there are no simple short-cuts regarding its derivation \cite{mvp}. This idea of MVP is close related with the inbreeding tolerance or preference for the species in order to survive. In fact it is very interesting that a small and isolated number of individuals could also could lead in few generation to a new species. Recently the first example of speciation in a very small isolated population of birds was observed directly in the field \cite{speciation}.

Based on the biological evidence of incest being tolerated or even preferred for some species, we considered some degree of incest along biparental species as an important key to build a general model. Using a Markov process, we obtained a more realistic tree of ancestors \cite{nuestro-paper}.

\subsection{On the paper structure}

We present a description of the problem through an algorithmic and analytic approach. The structure of this manuscript is developed as follows: Section \ref{build} is a description of the algorithm implemented to generate the trees of ancestors. Section \ref{rebuild} shows an equivalent description and a generalization in terms of mathematical symbols. Comparative results between the implemented algorithm and the theoretical model from \cite{nuestro-paper} are presented in Section \ref{compare}. Finally, in Section V we present conclusions and further work.

\section{BUILDING THE COMPUTATIONAL MODEL}\label{build}

\subsection{A simplified tree of ancestors} \label{sub-sec-1}

Let's start by creating a simplified tree of ancestors for one individual. We considered an index $n$ to label the generation number, starting from $n=0$ the \textit{parents} generation, $n=1$ the \textit{grandparents} generation and so forth. Initially we fixed the number of ancestors for the first generation, $n=0$, at $2$, because the individuals have sexual reproduction. In successive generations $n\geq1$ we run a random number generator, denoted by $r_n$, to obtain a number between $2$ and a maximum value that will depend on generation number. In each generation the maximum possible number of ancestors is constraint to:
\begin{equation}\label{equ_1}
R_n=2^{n+1}.
\end{equation}

To avoid a high endogamy degree at the beginning of the tree, we also fix this value at generation $n=1$ in $4$. Therefore, in the next generation, $n=2$, the maximum possible ancestor number is $R_2=8$. Now at $n=2$ for the first time we run a random generator using an uniform distribution to obtain a number between 2 and $R_2$. The possible maximum number for ancestors in the following generation, $n=3$ is less or equal to $R_3$. Actually, this value depends on the random value obtained in the generation before. 

In general, for one individual at the generation $n$ is not possible to have more ancestors that the double of the ancestors obtained in the previous one $n-1$ (two parents per each predecessor), i.e: 

\begin{equation}\label{cota superior mejorada}
r_{n}\leq 2 r_{n-1},
\end{equation}

and of course we have $r_n\leq R_{n}$, for all $n$. The expression \eqref{cota superior mejorada} is a \textit{better} upper bound that $r_n\leq R_{n}$, because $2r_{n-1}\leq R_{n}$.

\begin{figure}[!htbp]
\begin{center}
\includegraphics[totalheight=10cm]{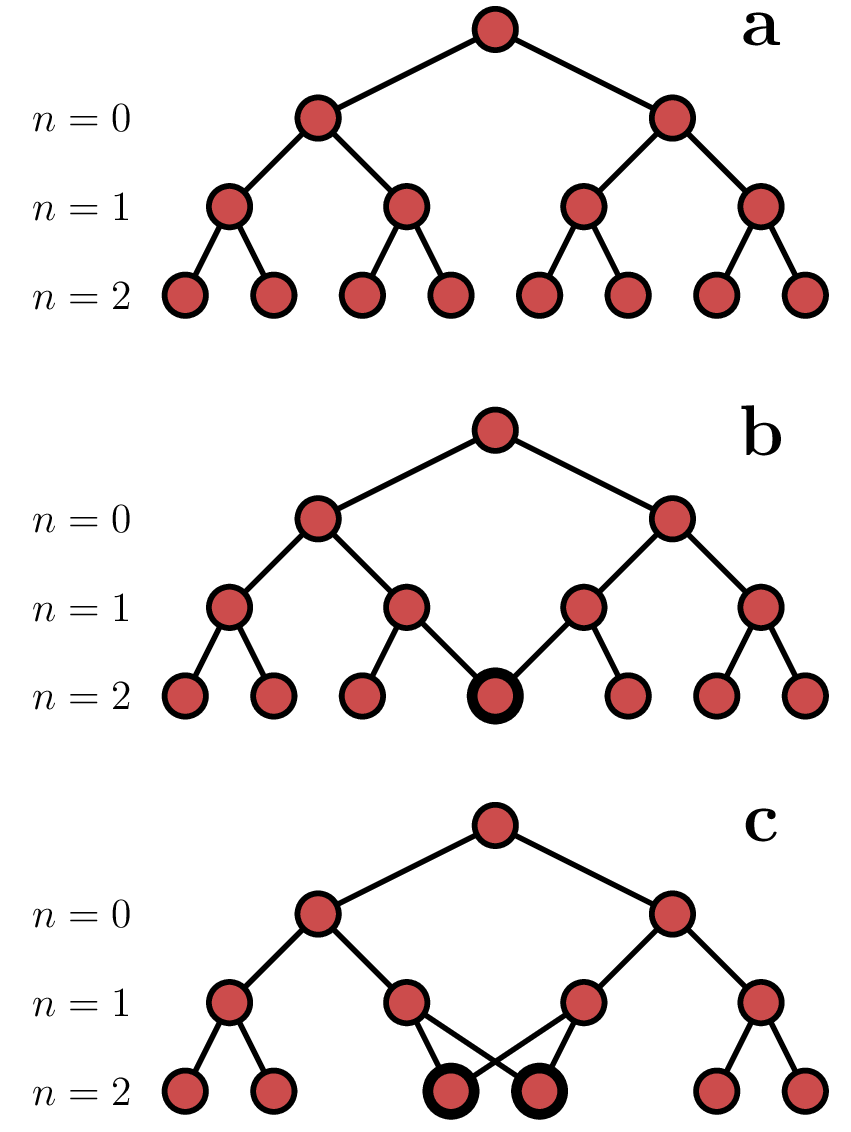}
\caption{(Color online). Example of three kinds of inbreeding paths for genealogical trees in three generations. First $(\mathbf{a})$ panel shows no restrictions by blood relationship. Second and Third panel $(\mathbf{b,c})$ shows two kinds of restriction in third generation: ancestors sharing one parent $(\mathbf{b})$ and ancestors sharing two parents $(\mathbf{c})$. The restriction by blood relationship increases according to the degree of endogamy as in \cite{nuestro-paper}.}
\label{fig:01}
\end{center}
\end{figure}

To illustrate a set of possible trees, the Figure \ref{fig:01} shows three simple examples for the first three generations in the chain.

For each $n-$generation in the tree, we obtain a random number $r_n$ between the limits given for the random number obtained in the previous generation, i.e. $r_n\in[2,2r_{n-1}]$, which accounts the degree of blood relationship, or endogamy, between the individuals of the same generation. A full tree is developed using a generation number loop increasing program. In order to avoid any possible bias in the tree we change the seed of the random generator in each run. Additionally we have established a generation in which the number of ancestors is maximum (more detail are presented in Section \ref{parameters-sec}).

\begin{center}
\begin{figure}[!htbp]
\includegraphics[totalheight=4.4cm]{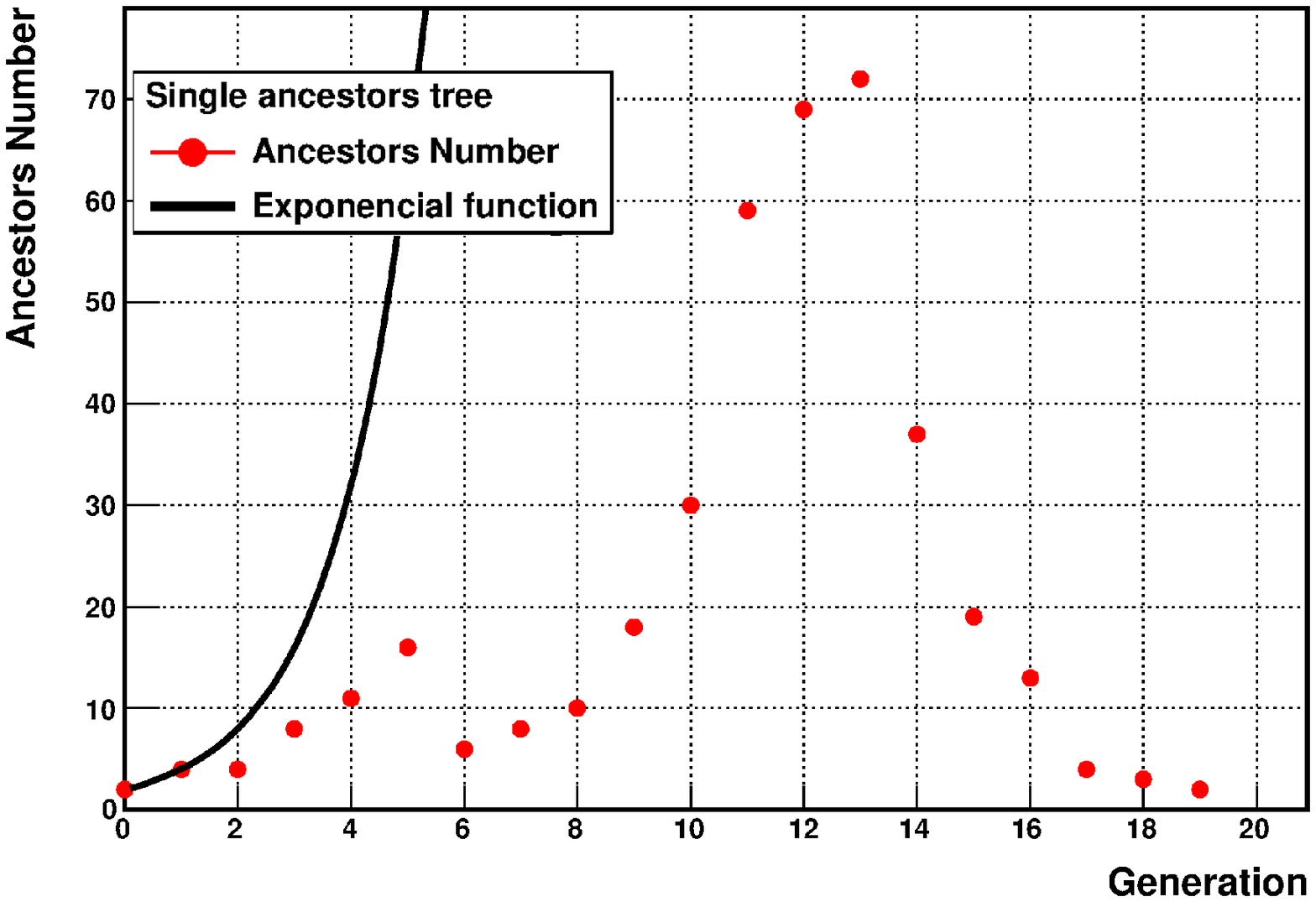}
\includegraphics[totalheight=4.4cm]{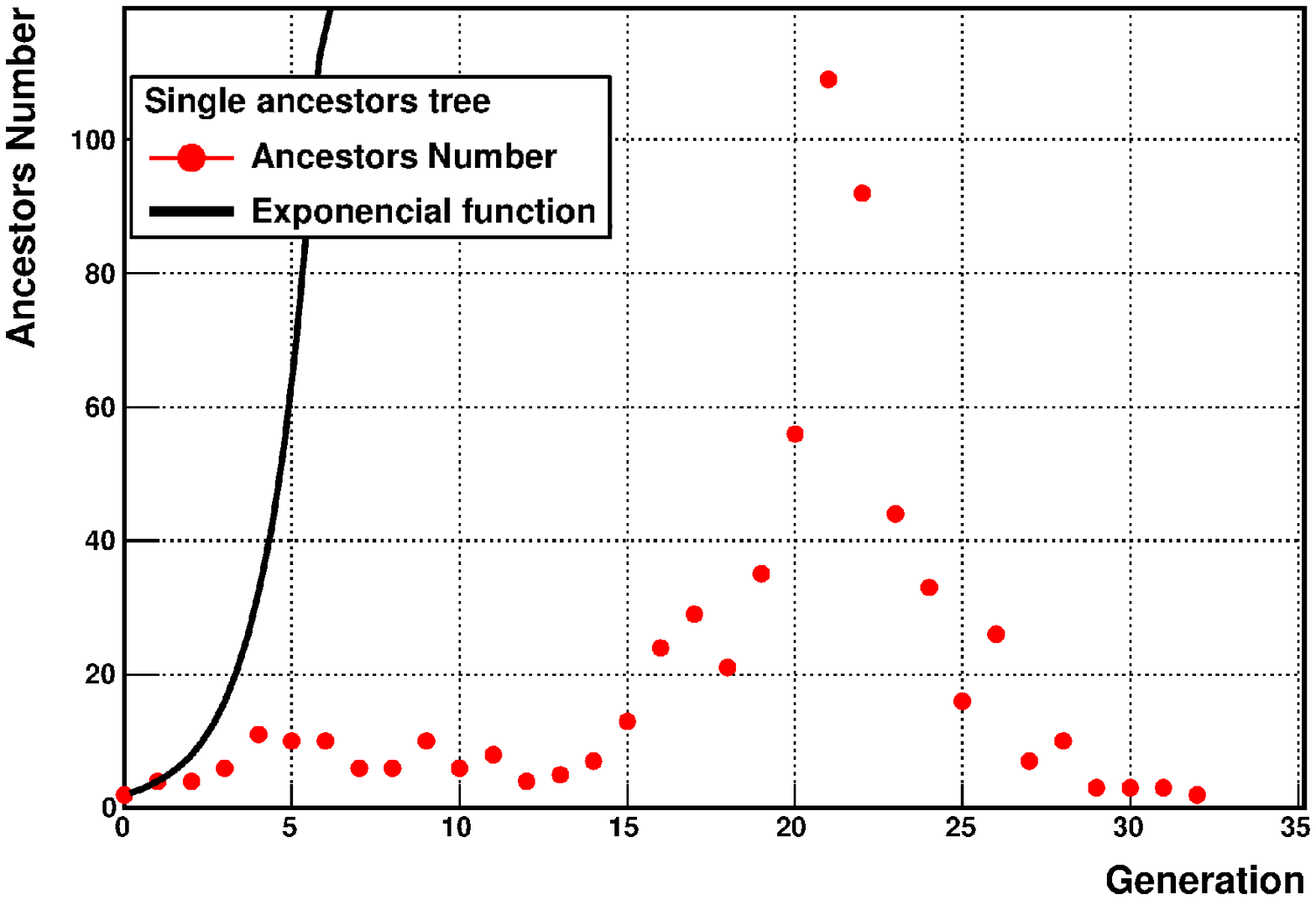}
\includegraphics[totalheight=4.4cm]{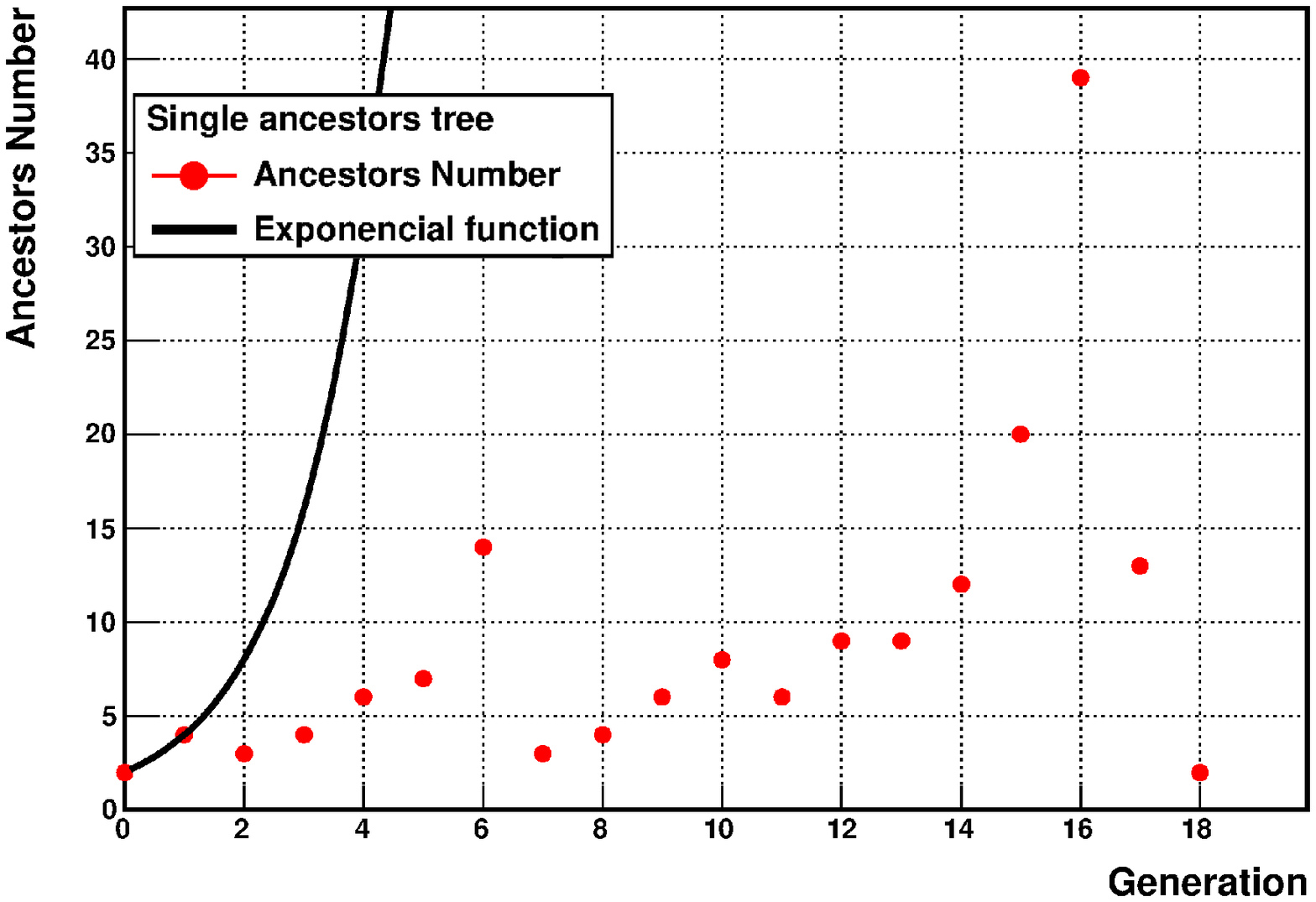}
\includegraphics[totalheight=4.4cm]{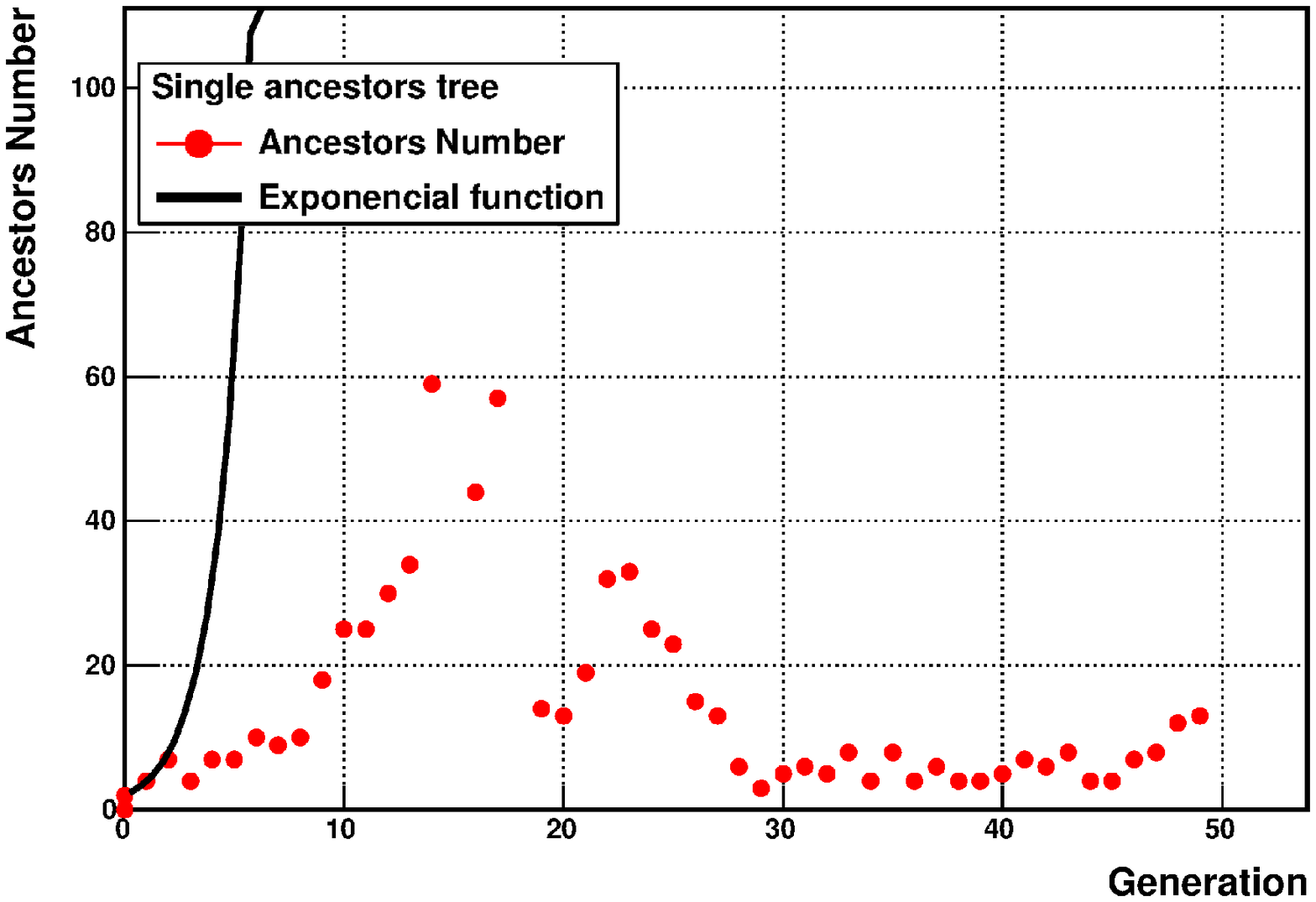}
\caption{\footnotesize{(Color online). Example of four kinds of genealogical trees from present time or generation to past. Red dots are the number of ancestors for each $n-$generation at most equal to $N=50$. All examples are compared with no-inbreeding tree, $R_n=2^{n+1}$, represented with the black line.}}
\label{fig:02}
\end{figure}
\end{center}

The length of the genealogical tree is as long as we prefer to define. For instance in our illustrative example we used $N=50$ generations as the absolute maximum. Actually this value depends on when the random generator reaches a minimum of $2$ ancestors, because at that generation the process will end. An important consideration is that the simulation of one particular tree could end earlier than the absolute maximum generation if, for a given generation, the number of ancestors reaches by chance the number 2. Four examples of random trees generated are shown in Figure \ref{fig:02}, where $R_n$ in \eqref{equ_1} corresponds to the case in which all ancestors are different in each $n-$generation. 

With all these considerations, very endogamous trees has been occasionally obtained. The explanation for this is that any possible combination of inbreeding between ancestors has the same probability (uniform distribution) and becomes very unlikely a non-inbreeding case. Therefore we need to introduce additional considerations over the endogamy degree. We improved the algorithm in order to provide a more realistic simulation.

\subsection{A more sophisticated tree of ancestors}\label{sofisti}

We found a way to have more control of the endogamy degree of the tree. The algorithm that we used has been enriched by endowing it with the possibility to switch the reproduction between inbreeding or not. A second random number generator was used in each generation $n$, namely $s_n$, that returns an integer $s_n=0$ or $s_n=1$.

\begin{figure}[!htbp]
\begin{center}
\includegraphics[totalheight=2.45cm]{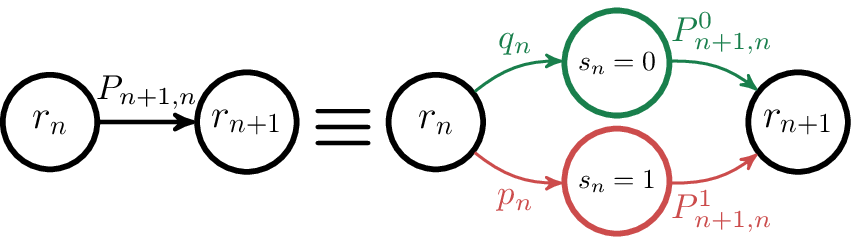}
\end{center}

\caption{\footnotesize{(Color online) Diagram of a generic link associated to the transition $r_n\longmapsto r_{n+1}$. There is $2$ branches from the intermediate state $s_n$, $q_n$ is equal to the probability to stay in $s_n=0$, upper branch (green), i.e. $q_n=P(s_n=0)$, $p_n$ is equal to the probability to stay in $s_n=1$, lower branch (red), i.e. $q_n=P(s_n=0)$ and $p_n=1-q_n$. The notation for the conditional probabilities is simplifed  $P^0_{n+1,n}=P(r_{n+1}|r_n,s_n=0)$ and $P^1_{n+1,n}=P(r_{n+1}|r_n,s_n=1)$.}}\label{diag1}
\end{figure}

The outcome of $s_n$ and $r_n$ can affect to $r_{n+1}$ stochastically. One of the leading actors in this algorithm is the distribution used for the random generator number $r_{n+1}$, taking into account the previous result for $r_n$ and the branch from which it comes. Analytically this is equivalent to a conditional probability for $r_{n+1}$ given $r_n$ and $s_n$: $P(r_{n+1}|r_n,s_n)$. A schematic diagram of the above description is shown in Figure \ref{diag1}. $s_n$ is binary random variable used to select between all possible cases divided in 2 types, at each generation.

On the other hand, we considered equal probabilities for each branch, i.e. $s_n=0$ or $s_n=1$: $P(s_n=0)=P(s_n=1)$. Then, of course, we have $50\%$ chances to obtain $s_n=0$, also the same chances for $s_n=1$, for each $n$.

Different reproductive behavior such as one male mating with several females in a group, could be changed via the percent rate in this second generator. Other behavior like sibling selection could be included via the probability distribution of animal preferences using another distribution of probability different of the uniform, to give different weights within the inbreeding selection option, between $2$ and $2 r_n-1$.

We explored different distributions of random numbers for the development of each individual tree, but we selected those that could be linked experimentally with an animal reproductive behavior. In this case we study a uniform distribution and a negative exponential distribution for the growth of the tree and a negative exponential distribution for the decay of the tree \cite{dist-idea-1,dist-idea-2}. More details will be exposed in the following section. 

\subsection{Details on the tree simulation parameters} \label{parameters-sec}

Due to the freedom in the design of the algorithm, as well as in its parameters, it is necessary to specify the information that leads to one possible development of the tree. Taking into account the experimental data, most of our assumptions could be replaced by constraints of animal nature. Our intention is that others could be able to use the algorithm, i.e. choose the parameters and the probability distribution, in a way that would be useful to represent real data sets. In this section we present our selection of parameters.
  
Each simulated tree has a maximum length of $N$ generations, but a given execution may finish before, as we explain in Section \ref{sub-sec-1}. The simulation process is mainly separated into two regimens: \textit{growth} and \textit{decay}; where the number of ancestors increases and decreases, on average, respectively. As we explained above in Section \ref{sub-sec-1} we chose a generation where the number of ancestors is maximum. We have not chosen the maximum number of ancestors, we have only chosen where this maximum is reached, we denoted this value by $N_\mu$. After this generation the number of ancestor could never exceed the random value obtained for $N_\mu-$generation. This generation defines the \textit{growth interval}: $[0,N_\mu]$. In this paper we used $N_\mu=N/2$. The \textit{decay interval} is defined from $N_\mu$ to $N$. 

Both parts of the tree development, \textit{growth} and \textit{decay}, are ruled almost by the same algorithm: a second random number generator $ s_n $ is used to make a bifurcation rule. All these bifurcations lead to different actions depending on the regime we are dealing with, i.e. growth or decay. 

For the \textit{growth} regime (the interval $[0,N_\mu]$), if $s_n=0$, the number of ancestors at generation $n+1$ will be the maximum possible with no inbreeding: $r_{n+1}=2 r_{n}$. This means that each ancestor in the generation $n$ has all different parents in the generation $n+1$, i.e. the pairs of parents of each individual of generation $n+1$ are different one to one. If $s_n=1$, we will have for $r_{n+1}$ any kind of inbreeding reflected between $2$ and $2 r_n-1$. For this branch we used a uniform or a negative exponential distribution for the random number $r_{n+1}$.

The \textit{decay} regime is subdivided in two parts, the first one is defined from $N_\mu$ up to a cut generation $N_\gamma$. In this interval, $[N_\mu,N_\gamma]$, the branches $s_n=0$ and $s_n=1$ are equally the same as in the growth regime, the only difference here is that a negative exponential distribution is used for the random number $r_{n+1}$.

The second one is a \textit{harder decay} version of the first part.  In the interval $[N_\gamma,N]$ and branch $s_n=0$, the not endogamy rule is replaced by a full-endogamy rule.

We have defined full-endogamy as the case where the number of ancestors is fixed and constraint to the number of individuals at the beginning of the ancestors tree. In an extreme case, the minimal number of individuals prefixed would be 2 or 4, but could be any number. This pre-fixed number represents the upper bound over the initial number of individuals at the beginning of the population who originated the tree. In other words, we have merged random endogamy probability with the probability of having a low fix number of ancestors; this, represents the upper bound of individuals at the beginning of an initial population. The $s_n=1$ branch remains the same, the only change is that it uses a negative exponential distribution for the random number $r_{n+1}$. 

For this reason, we call the first part as \textit{soft decay} and the second part as \textit{hard decay}. 

In all Figures presented in this work, the value of $N_\gamma$ is selected 10 generations before $N$. Even when some trees could end by chance before this maximum generation, we used this rule to conduce to converge to the maximum possible value of individuals who originated the tree, at $N-$generation could be 2, 4 or any other value. The number of ancestors at the end of tree is constrained to the number of individuals who originated a particular population.

To summarize, the development of tree could be split in three steps delimited by this four generations ordered as follows: $0<N_\mu<N_\gamma<N$. The \textit{growth} region starting from generation $0$ to $N_\mu$ and for the \textit{decay} region consisting in two intervals: \textit{soft decay} from $N_\mu$ to $N_\gamma$ and \textit{hard decay} from $N_\gamma$ to $N$.
  
We could add in the simulation a condition to request that not all possible inbreeding options be equally likely. To do that we used different distribution in the growth to represent differences in animal mate preferences. We compared a simulation using the uniform distribution, that corresponds to no preferences in mate selection, with a simulation using the negative exponential distribution (siblings preference). In the case of the second option, there is an additional free parameter in the distribution that allows us to control the endogamy degree.

The exponential decay case corresponds to a very inbred preference of the ancestors (close kin), where an increasing exponential distribution corresponds to avoid the incest as long as possible (within the random case). The opposite case is an increasing exponential distributions. It corresponds to a preference selection of a not close kin when selecting a kin for mate.

We have chosen a \textit{shift} in the negative exponential distribution that depends on the generation. This shift was chosen in order to obtain the maximum endogamous probability at the half of the interval between $2$ and $2r_{n-1}-1$ for each generation number $n$. Regarding the decay part, our selection was to use always the exponential decay distribution, since we want to give in the decay  more weight to the endogamous preference.

Other additional distributions could be used, such a gaussian distribution. In this case there is an additional parameter to fix. Those parameters does not necessary represent an aspect in animal mate selection. We did not use such distribution because we want that all parameters used represent an aspect on the animal mate preference. Nevertheless, the gaussian distribution could be used in other similar simulations based on the ancestors simulation algorithm to represent some other markovian process. 

In Section \ref{rebuild} we show a way to describe the simulation in terms of a theoretical stochastic process. That allows to describe the expected value of $r_{n+1}$ given the values of $(r_n,s_n)$, as a recurrence equation, using the \textit{Law of total probability}. The expected value for $r_{n+1}$ depends on two terms that corresponds to the cases $s_n=0$ or $s_n=1$.

\subsection{Building an ensemble of trees}

Once proposed the rules to generate the tree for one individual and implemented it, we studied what happens with a general population with independent ancestors trees generated in such way (Section \ref{parameters-sec}). The process consists of generate N independent random trees for a set of not related individuals and then take the average of the total population, generation by generation. In that way we have a representative mean behavior of the trees population for each generation.

The average of the trees samples could be compared with the expected value of the random variable in the first theoretical model \cite{nuestro-paper}.

Our first trial to study a set of trees was to consider a population of 50 individuals (each one with its own tree). The set is large enough to perform the statistical analysis but at the same time no so much to reduce computational time. In this way we can study the different effects on the selection of model parameters, i.e. we averaged each generation of the independent 50 trees.

There are other ways to combine the trees without using the mean of the sample set, for instance by means of a genetic algorithm \cite{gen-1}. The arbitrary choice of a parameter over others to optimize the trees (fitness) and use them as seeds to generate new trees prevents us to follow this approach. Additional studies with data from the biology field are necessary to follow the genetic algorithm approach. 

Any particular selection made in the model is pointed out in order to be clear for the further reader where to replace it with one according to a specific group of animals. Then with the replacement it is possible to re do the calculation of the mean value of ancestors number.

\subsection{Results of tree sets}\label{sec-dist}

We used different distributions for the random number generator as described in Section \ref{parameters-sec}. The uncertainties in all Figures are the bin errors associated with each $y_b$, the data of the $y-axis$ in the $b-$bin. This quantity is calculated as the standard error on the mean as $\Delta y_b = S(y_b)/\sqrt{n_b}$, where $S(y_b)$ is the standard deviation $(RMS)$ of the $y_b$ data and $n_b$ is the number of bin entries. In this case the entries are the number of trees that are considered in the calculation of the mean at the $b-$bin. 

\begin{center}
\begin{figure}[!htbp]
\vspace{0.1cm}
\includegraphics[totalheight=4.46cm]{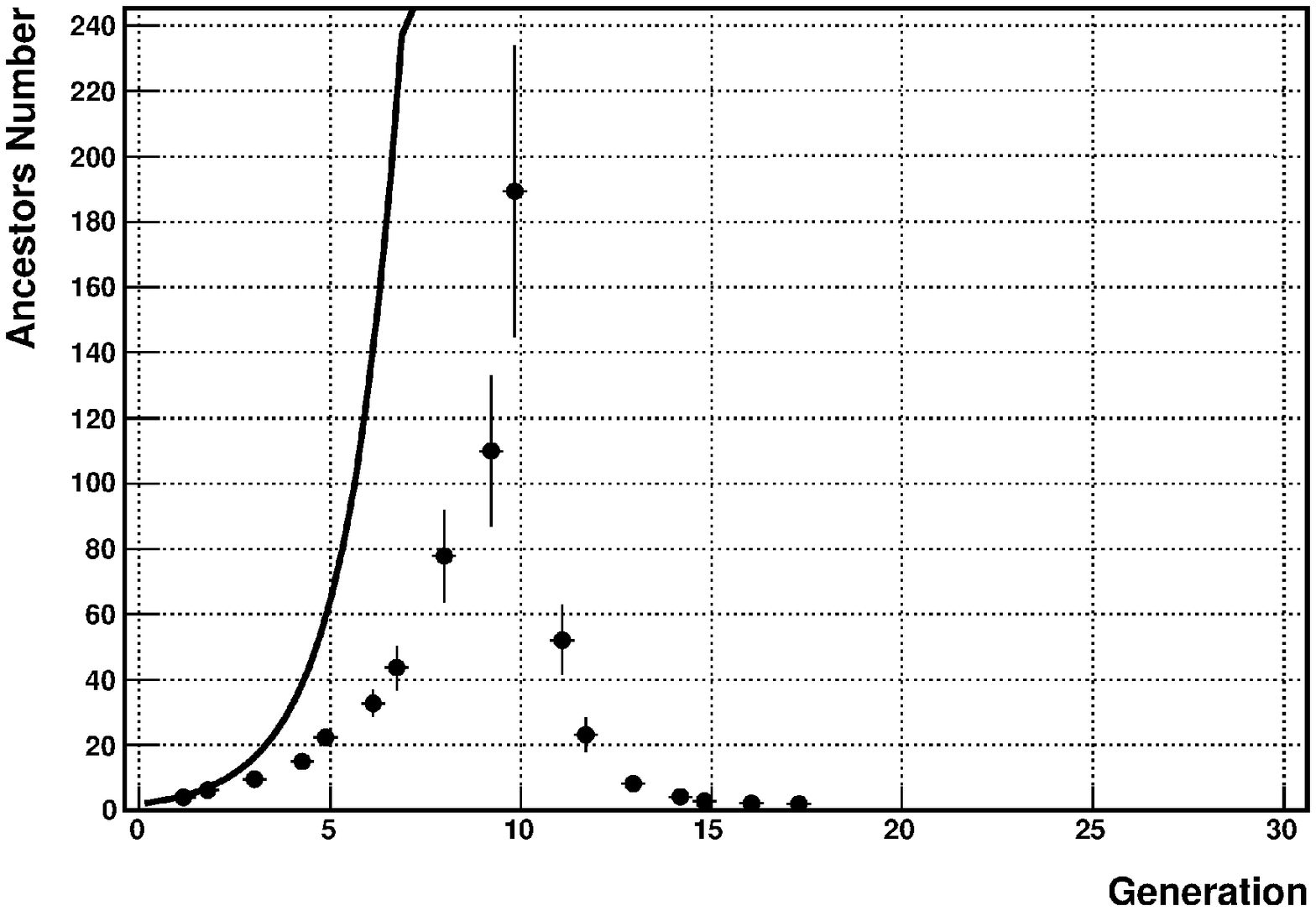}
\includegraphics[totalheight=4.46cm]{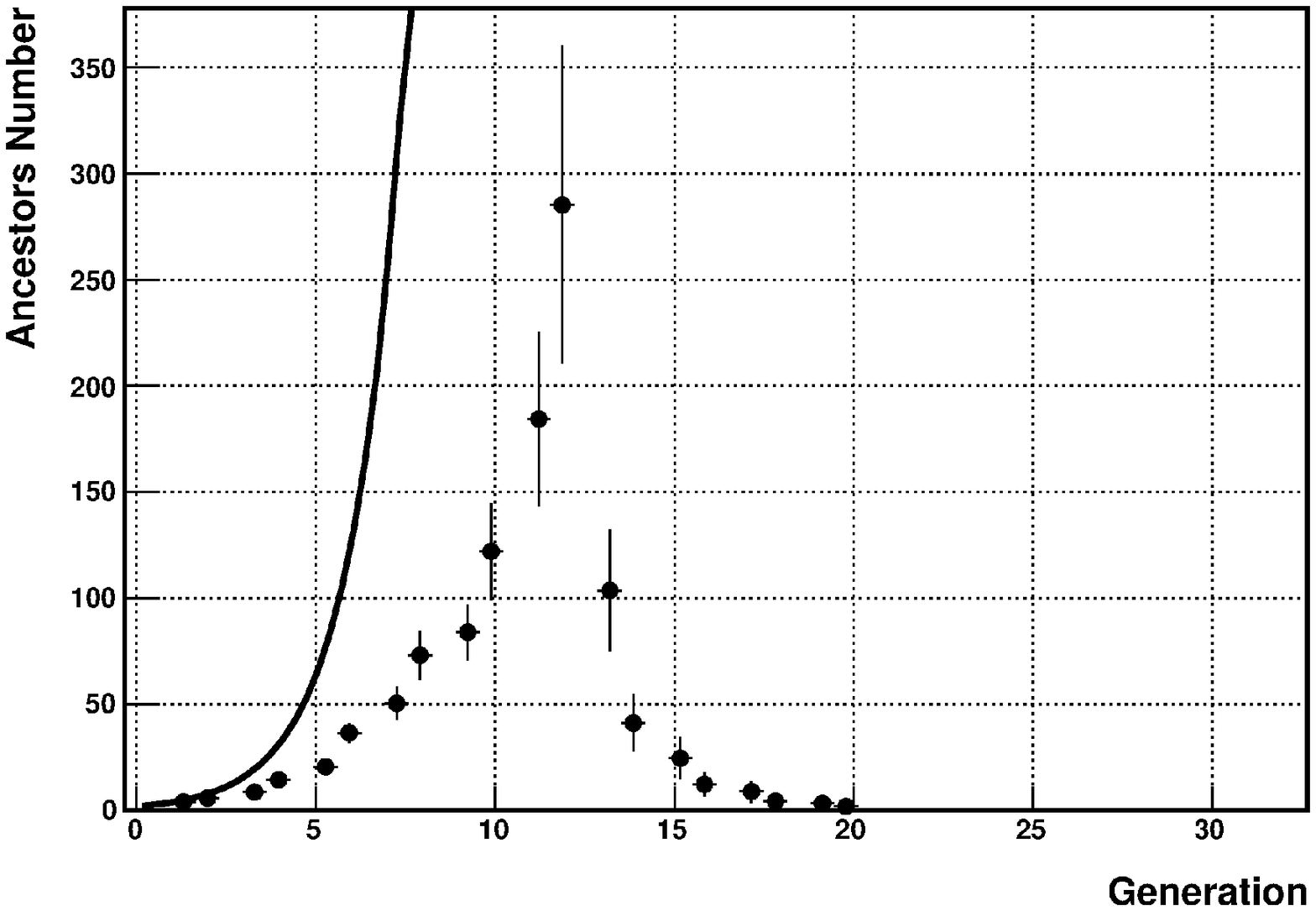}
\hspace{0.2cm}
\includegraphics[totalheight=4.46cm]{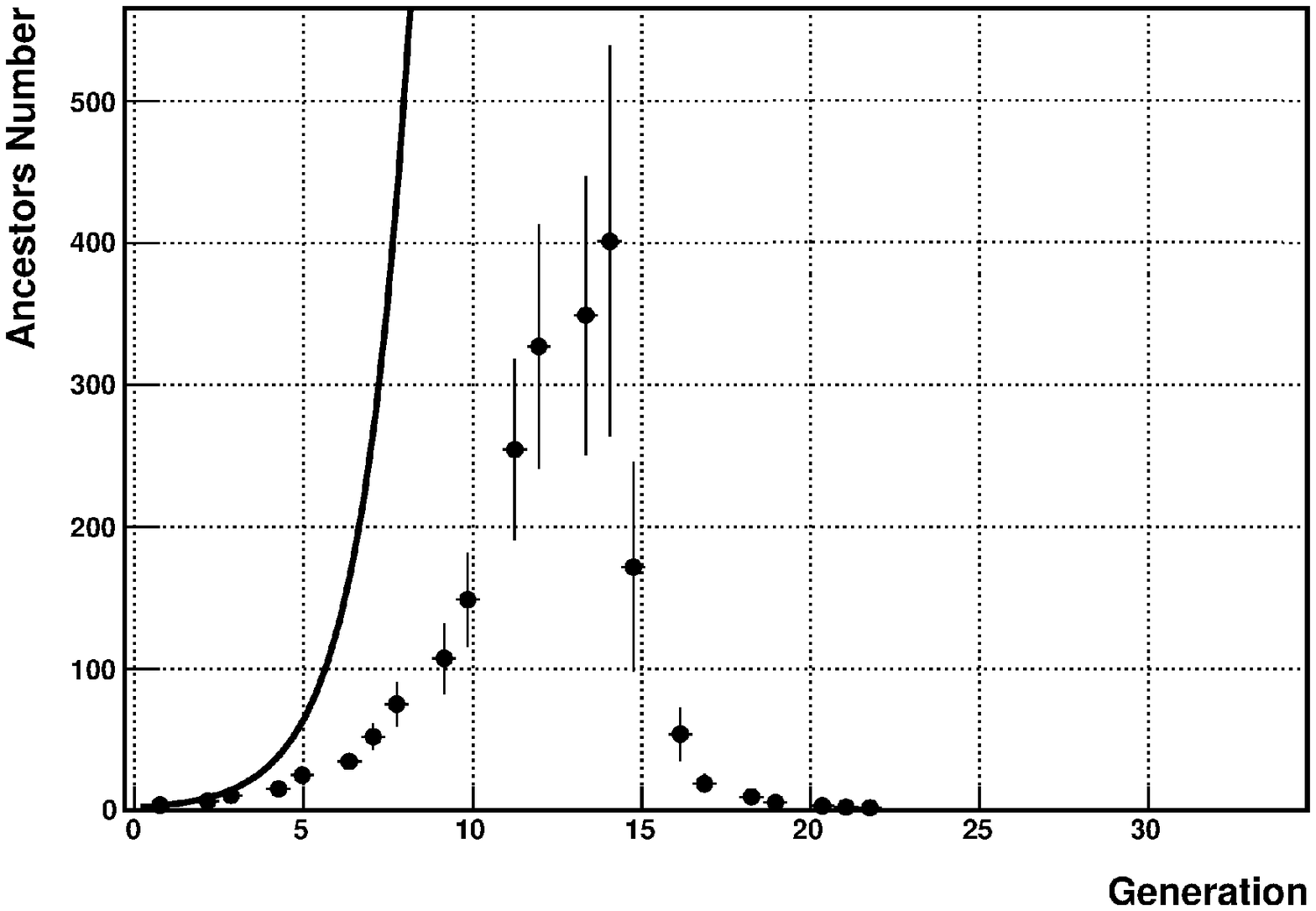}
\includegraphics[totalheight=4.46cm]{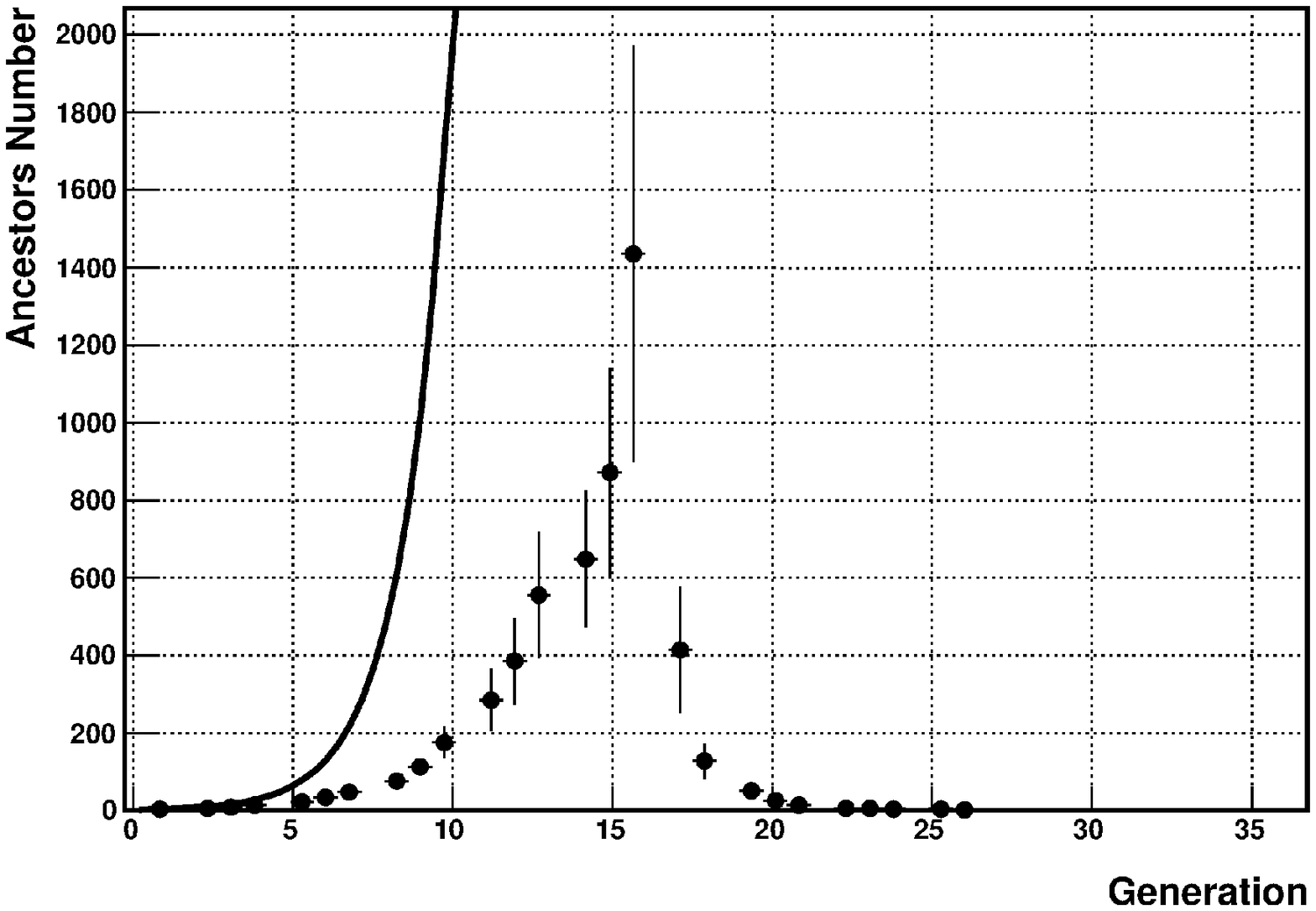}
\hspace{0.2cm}
\includegraphics[totalheight=4.46cm]{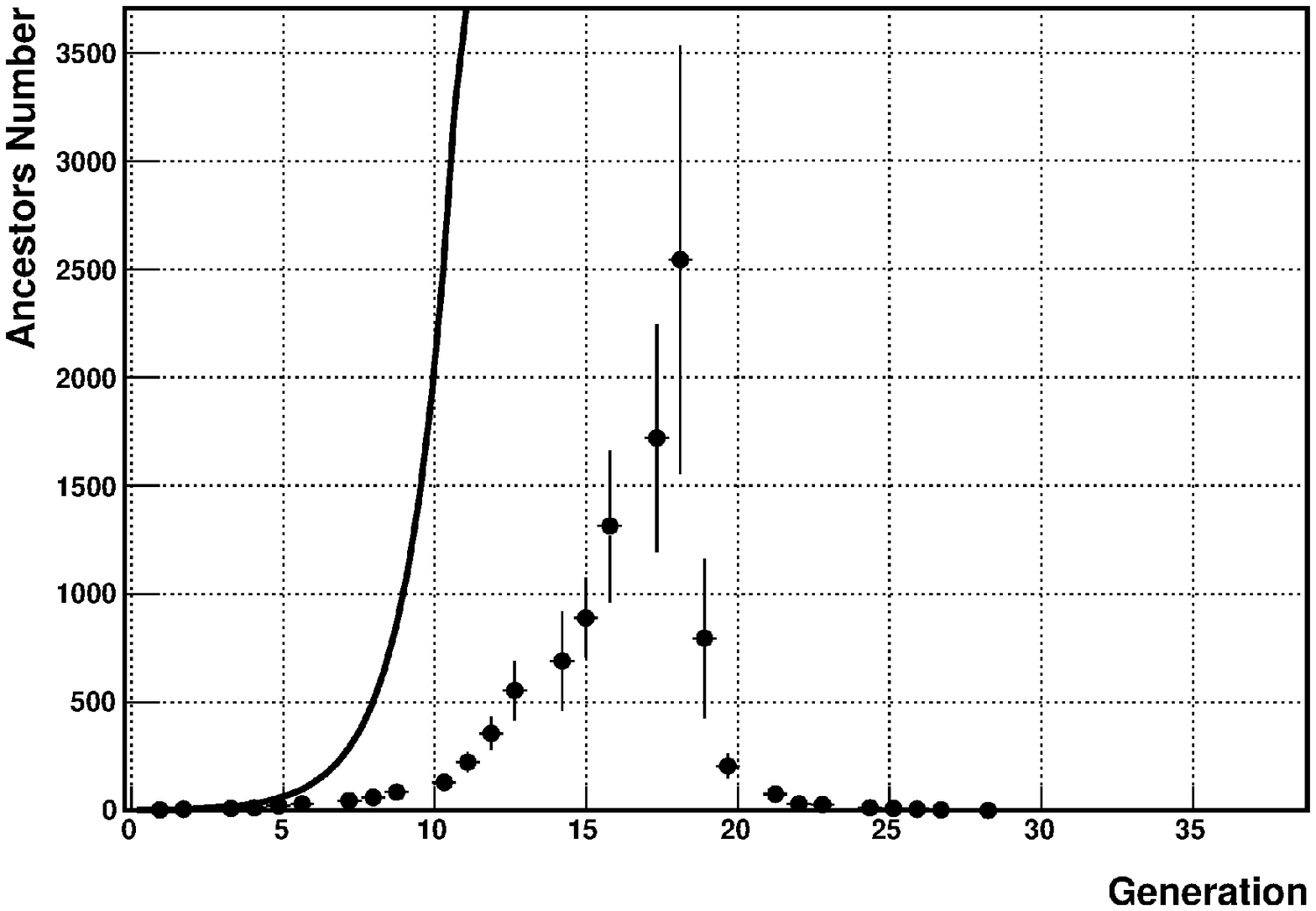}
\includegraphics[totalheight=4.46cm]{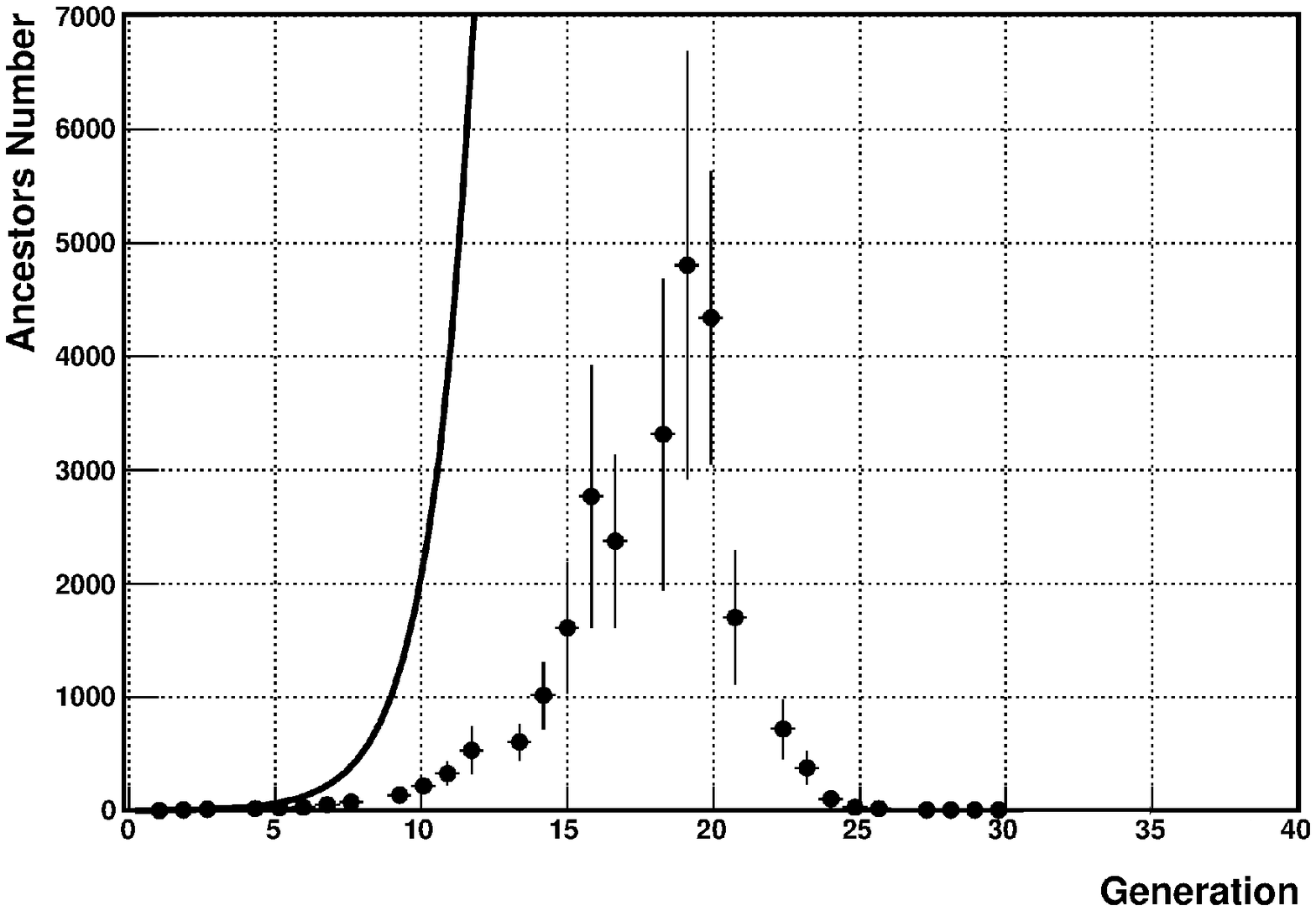}
\vspace{0.2cm}
\caption{Mean number of ancestors for each generation. Average of 50 trees. For each tree the random generator uses an uniform distribution of random numbers. From left to right and  from top to bottom the maximum generation in each case is $N=20$, $N=22$, $N=24$, $N=26$, $N=28$ and $N=30$.}
\label{fig:03}
\end{figure}
\end{center}

The simplest case consist of taking the uniform distribution. The results obtained with uniform distribution are presented in Figure \ref{fig:03} (right).

In Figure \ref{fig:03} are shown different selection for the generation where the number of ancestors starts to decrease and a maximum generation to end the process, $N$, previously described in \ref{parameters-sec}. In each case, the mean number of ancestors grows, reaches a maximum value and then decreases, depending on the maximum generation to end the process. 

Figures show process ending up to generation $N=20$, increasing this value up to $N=30$. When the maximum generation is changed, the evolution of the mean value is affected as it is shown in Figure \ref{fig:03}. 

\begin{figure}[!htbp]
\begin{center}

\hspace{1cm}\includegraphics[totalheight=6cm]{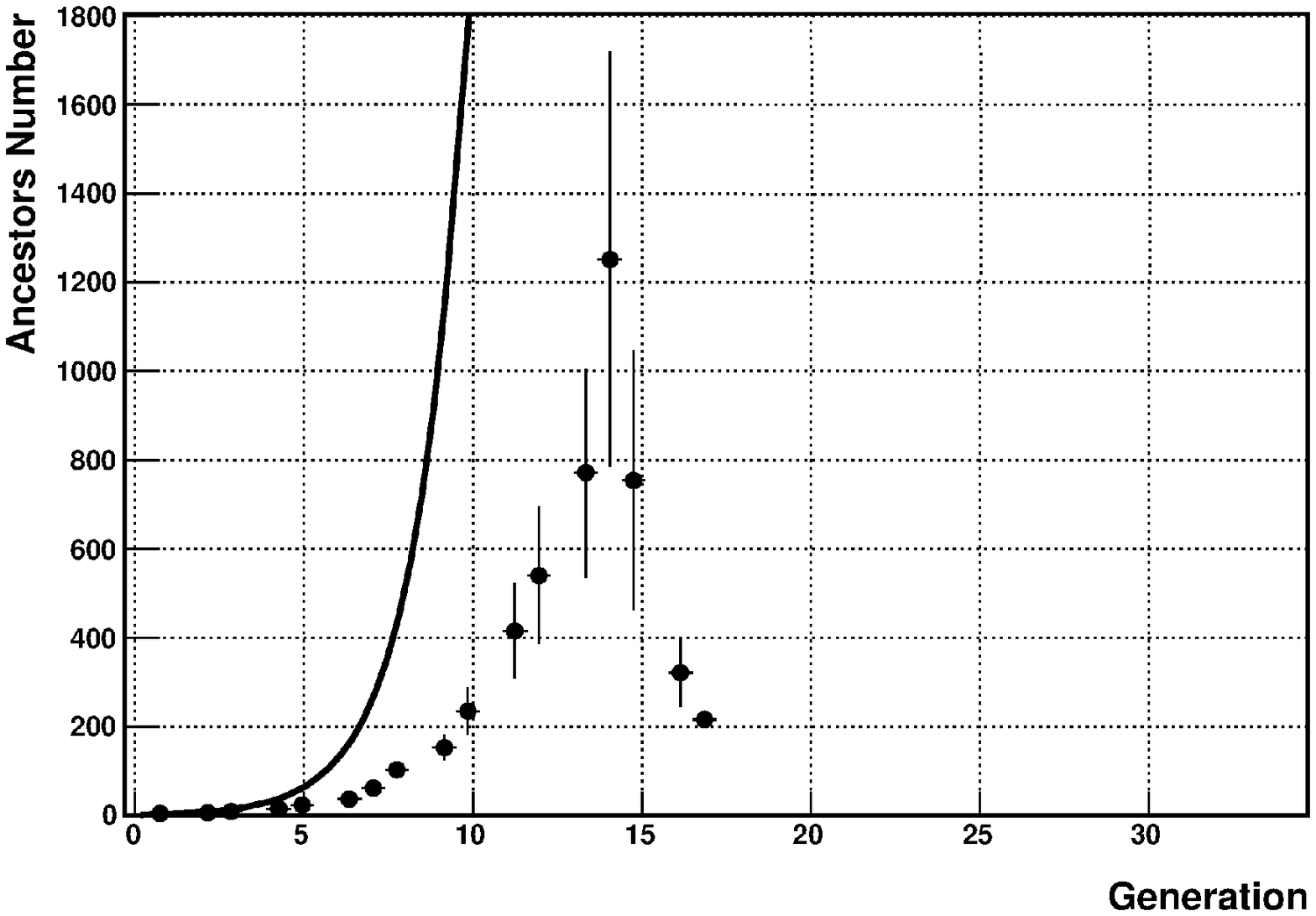}\\
\hspace{1cm}\includegraphics[totalheight=6cm]{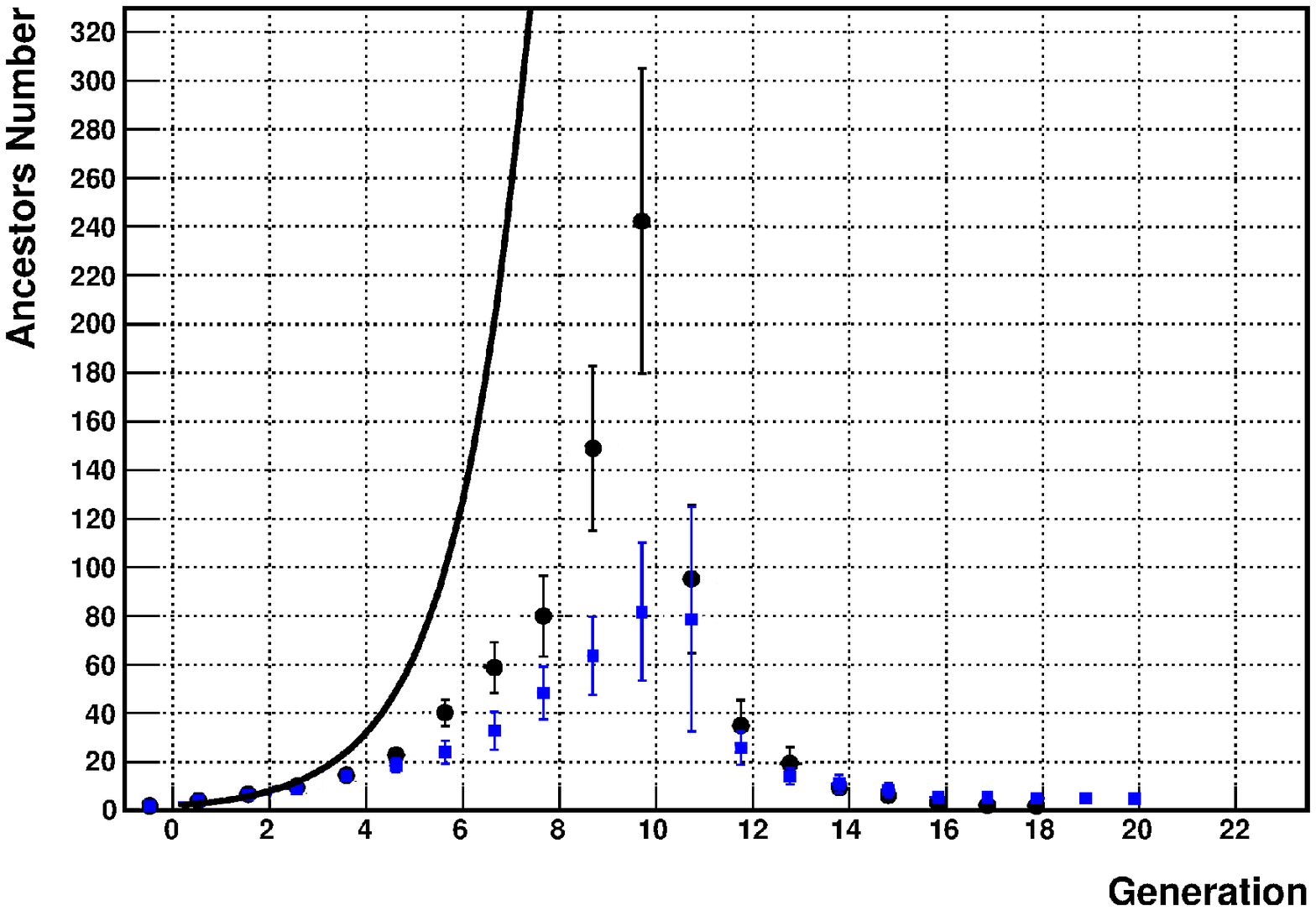}
\caption{\small (Color online) TOP: Mean number of ancestors for each generation. Average of 50 trees starting with an initial population of 100 pairs of animals, with uniform distribution. BOTTOM: A comparison for the 50 ancestors tree generated with uniform distribution (above) and negative exponential distribution bellow.}
\label{fig:04}
\end{center}
\end{figure}

Other possible setting for the tree of ancestors consist of do not finish the process when the number of ancestors is two and to end it with a different number of ancestors as initial population as we explained in Section \ref{parameters-sec}. This case corresponds to a particular population starting not with 2 original ancestors, but with other number of pairs of animals, such as 200 original individuals as the initial population shown in Figure \ref{fig:04}. Initial constraint on population is a parameter that can be changed in order to compare with different animal population under study. 

A comparison between the case using the uniform distribution and the exponential decade distribution is shown in Figure \ref{fig:04}. The average of 50 trees generated with each one are shown with dots and squares. As we expected, smaller values for the number of ancestors are obtained when using the exponential distribution for the same values of initial population and maximum generation. 

Figure \ref{fig:04} (Top panel) corresponds to the case where the initial number of ancestors is different to 2, using the uniform distribution. The Figure \ref{fig:04} (Bottom panel) corresponds to the comparison between the uniform distribution and the negative exponential distribution.

Regarding the simulation, the code execution collects a certain number of realizations, i.e. a number of generated trees $N_T$. In $n-$generation we compute the number of trees, coming from executions of the code, that have not finished in that generation, whose result for $r_n$ is \textit{equal to} $r$, denoted by $T(n|r)$, divided by the total number of trees, coming from executions that have not finished in that generation also, denoted by $T(n)$. This quotient could work as a pdf of $r_n$ 
\begin{equation}\label{quotient}
\mathbb{P}[r_n=r]\;\mathlarger{\thicksim}\;\frac{T(n|r)}{T(n)},
\end{equation}
the symbol $\mathlarger{\thicksim}$  represents a  more properly correspondence satisfied when $N_T$ grows, i.e. a large number of code executions. Each generated tree corresponds to a certain histogram, distributed in mutually exclusive cases, $r=2,3\,\cdots$, then the quotient on the right side of \eqref{quotient} will be properly normalized following:

\begin{equation}
\mathlarger{\sum}_{r} \,\frac{T(n|r)}{T(n)}=1.
\end{equation}

If we have from the simulation a quantity \textit{almost} equal to the $r_n$ distribution, we can obtain the expected value, or even more, the higher $k-$order moment for each generation $n$, denoted by $\langle r_n^k\rangle$, associated to the distribution using the definition provided by the theory of probability 
\begin{equation}
\langle r_n^k\rangle\,\mathlarger{\thicksim}\,\mathlarger{\sum}_r \;r^{k}\,\frac{T(n|r)}{T(n)}.
\end{equation}

With this simulation we built individual trees and ensembles of trees for animals with sexual reproduction. We can use them to show some characteristics of the process. In the simulation we represented the animal preference for a mate by a set of free parameter of the proposed algorithmic model. The examples that we presented are not intended to be exhaustive, although these examples could give an idea about the results of parameter selection. The preference could be taken from experimental data, with a different number of individuals of the population under study or different group of original ancestors. Further models could include other characteristics of particular animal groups, even a different way to generate the ensemble of trees.

\subsection{Regarding model computation of a set of trees}

Another aspect to consider is the time that takes to create a full tree. This variable depends on the length of the tree, longer trees needs more computational time. 
Simulations up to 100 generations, when we generate $2\times10^{5}$ trees, takes much more than 4 hours. The current code implementation is linear with the generation number. It could be improved in the future using parallel programming tools \cite{charla}.

The model of the tree described here has been implemented in C++ using ROOT libraries \cite{root}. This object oriented framework has been used to analyze the results of the simulation too. All tools used in the work correspond to open source packages. Further implementations includes updates in the simulation that uses graph theory to study links or kinship between the parents \cite{ICTP-new}.

\section{REBUILDING THE COMPUTATIONAL MODEL ANALYTICALLY}\label{rebuild}

In this section we describe the simulation from the analytical point of view, considering a theoretical stochastic process. We present here the tools to describe the algorithm analytically. We can describe not only the trees of ancestors, we can also generalize the process, in order to study other problems.

For the trees described in Section \ref{sofisti}, we consider the stochastic process of two random variables $(r_n, s_n)$ on \textit{discrete time} $n$. As we showed before, the random variable $r_n$ represents the ancestor number at generation $n$. The role of the random variable $s_n$ is to define paths for the values that can take $r_{n+1}$, given the acquired value at previous generation, $r_n$.  We essentially distinguish the two random number generators used in the algorithm from the two random variables $ (r_n, s_n) $ associated with this analytical description of the algorithm. We used the same notation as the algorithm, by extension.

In Section \ref{sub-sec-1} we showed that for $s_n=1$ case the sample space of $r_n$, $\pmb{\mathtt{R}}_n$, is delimited by the obtained value for $r_{n-1}$, i.e. $\pmb{\mathtt{R}}_n=[2,2r_{n-1}]$. As a general rule, we \textit{throw} a random number, $s_n$, at generation $n$ between two possibilities with any labels, for instance $\{0,1\}$. According to the result of $s_n$, we choose one of the following two branches, where $\pmb{\mathtt{R}}^{s_n}_{n+1}$ is the set of values for $r_{n+1}$ given  $s_n=0,1$ and also $r_n$. The transition behavior of the process is governed, essentially, by the conditional probability $P(r_{n+1}|s_n,r_n)$ as we showed in Figure \ref{diag1}. In this paper we used different distribution according to the current subprocess. 

In Section \ref{sofisti} we introduced the \textit{growth} and \textit{decay} regimes, in this section we described them analytically as subprocesses of the main stochastic process. We can distinguish this two subprocesses in the following way: \\ 

$\blacktriangle$ \textit{Growth subprocess} $n\in[0,N_\mu]$\\

If $s_n=0$, then we take $r_{n+1}=2\,r_n$, i.e. we assign the double of the number of ancestors obtained in generation $n$, as the new number of ancestors for generation $n+1$. If $s_n=1$, we take $r_{n+1}$ as a random variable, with a \textit{certain probability distribution} whose support is $[2,2\,r_n-1]$.  We used two kind of distribution for $P(r_{n+1}|r_n,s_n=1)$ \textit{uniform} and \textit{negative exponential}.
\vspace{0.5cm}

$\blacktriangledown$ \textit{Decay subprocess} $n\in[N_\mu,N]$\\

The algorithm is similar, but is subdivided according to the generation number $n$:\\

\textbf{1.} \textit{Soft Decay} $n\in[N_\mu,N_\gamma]$. For the branch $s_n=0$, we continue with the use of  \textit{delta} distribution that assign for $r_{n+1}$ the double of the previous generation: $r_{n+1}=2r_n$. For the branch $s_n=1$, we used only \textit{negative exponential} distribution for $P(r_{n+1}|r_n,s_n=1)\sim e^{-\lambda_n r_{n+1}}$ over the set $\pmb{\mathtt{R}}^1_{n+1}$. The characteristic parameter of this distribution is $\lambda_n$, it is chosen in order to have more predominance to the smaller values of $\pmb{\mathtt{R}}^1_{n+1}$. Moreover, we add a constraint: the value of $r_{n+1}$ is bounded through the value obtained in $N_\mu$, $r_\mu$, that is $r_{n+1}\leq r_\mu$. \\

\textbf{2.} \textit{Hard Decay} $n\in[N_\gamma,N]$. The branch $s_n=0$ changes slightly. In this case we choose a random value uniformly distributed between $2$ and $\alpha_N$, in order to enforce the decay. This number $\alpha_N$ is pre-fixed and represents the upper bound of the initial number of​ individuals at the beginning of the population that has made the complete tree. The branch $s_n=1$ is ruled by the same rule as in the \textit{soft decay} regime. Also we still use $r_{n+1}\leq r_\mu$. 

We only use the \textit{negative exponential} distribution for $P(r_{n+1}|r_n,s_n)$. The value $r_{n+1}$ is bounded through the value obtained in $N_\mu$, $r_\mu$, this is  $r_{n+1}\leq r_\mu$.  

\begin{figure}[!htbp]
\begin{center}
\includegraphics[totalheight=2.5cm]{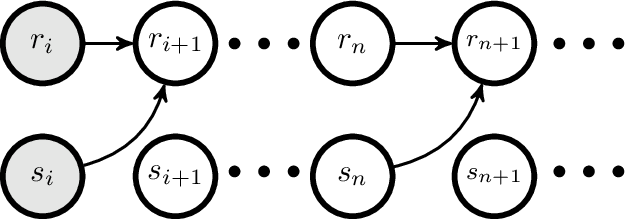}
\end{center}
\caption{\footnotesize{Diagram associated to the simulation process. The value of the random variable $r_{n+1}$ depends of the value obtained in the previous generation for $r_n$ and $s_n$. The node $r_i$ corresponds to the initial condition equal to $2^{i+1}$ at the initial generation $i=0,1$. The initial condition for $s_i$ is also a random number on $\{0,1\}$, not a prefixed number}}\label{diag2}
\end{figure}

A general diagram in Figure \ref{diag2} summarizes the sophistication implemented in the algorithm. 

In any regime we have
\begin{equation}\label{exclusive eq}
P(r_{n+1})=\sum_{s_n=0,1} P(r_{n+1},s_n)
\end{equation}
because the random variable $s_n$ defines a particular branch, which involves an \textit{exclusive} action over $r_{n+1}$.   

In the simulation proposed we have chosen the same probability distribution for all Bernoulli random variables  $\{s_n\}$: $P(s_n=0)=\frac{1}{2}$, for all generations $n\geq i$. This corresponds to the situation of maximum ignorance (disorder) regarding the process. We can generalize this situation denoting $p_n:=P(s_n=1)$ and the complementary probability $q_n:=P(s_n=0)=1-p_n$. Using the definition of conditional probability $P(r_{n+1},s_n)=P(r_{n+1}|s_n)P(s_n)$ and from equation \eqref{exclusive eq} we express $P(r_{n+1})$ as a convex combination 
\begin{equation}\label{exclusive eq2}
P(r_{n+1})=q_n P(r_{n+1}|s_n=0)+ p_n P(r_{n+1}|s_n=1)
\end{equation}



By definition, the expected value is
\begin{equation}\label{2}
E(r_{n+1}):=\underset{r_{n+1}\in \mathtt{R}_{n+1}}{\mathlarger{\sum}}\underset{}{} r_{n+1} P(r_{n+1}),
\end{equation}





The expression \eqref{exclusive eq2} allows to get an expression for the expected value $E(r_{n+1})$ from \eqref{2}
\begin{equation}\label{exclusive expected}
E(r_{n+1})=q_n\; E(r_{n+1}|s_n=0)+p_n \; E(r_{n+1}|s_n=1),
\end{equation}
this expression reveals the contributions from each branch to the expected value.

The simulation can be mathematically reformulated as follows: the random number obtained in generation $n+1$, $r_{n+1}$, is affected by the result of two random values in generation $n$: $(r_n,s_n)$. We say that in each generation the random variable $r_{n+1}$ run over a certain support $\pmb{\mathtt{R}}_{n+1}$ that depends of the values of $(r_n,s_n)$ at previous generation. 

The dependence of the random variable $r_{n+1}$ with $r_{n}$ is implicit in the equations \eqref{exclusive eq} and \eqref{exclusive expected}. It appears in the sample space $\pmb{\mathtt{R}}_{n+1}$ explicitly and in the parameters of $P(r_{n+1}|s_n)$ occasionally. Furthermore the diagram in Figure \ref{diag1}, represents a link in the process and it does not show the probability distribution $P(r_n)$ explicitly. Nevertheless, in order to provide a more specific description of transition $r_n\longmapsto r_{n+1}$ we need also to write the probability $P(r_{n+1})$ in terms of $P(r_n)$. The notation is simplified: $P(r_n)$ means the pdf associated to the random variable $r_n$. 

\begin{figure}[!htbp]
\begin{center}
\includegraphics[totalheight=5.5cm]{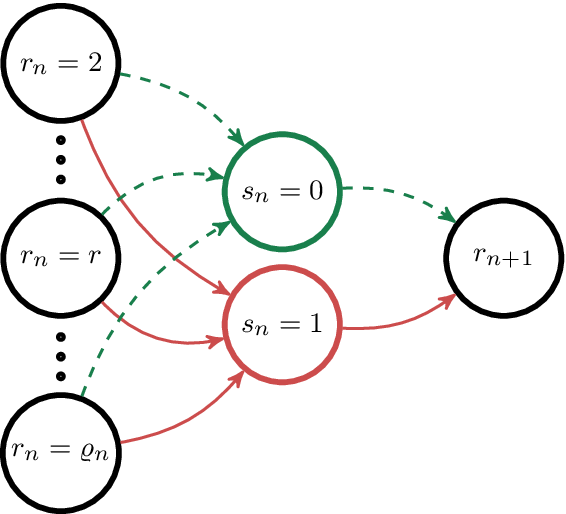}
\end{center}
\caption{\footnotesize{(Color online) On the left we show the specific values that $r_n$ can take, from $2$ to the maximum, denoted by $\varrho_n$, this value is related to the last link. Each value of $r_n$ has two possibilities (branches) to jump on the intermediate state $s_n$, i.e. $s_n=0$ dashed line (green) and $s_n=1$ continuous line (red).  
}}\label{diag3}
\end{figure}

As we show in Figure \ref{diag3}, given these two values $(r_n,s_n)$ there is a certain probability to jump on a particular value $r_{n+1}$. Then from equation \eqref{1} we have

\begin{align}\label{1}
P(r_{n+1})&=\mathlarger{\sum_{s_n=0,1}}\;\mathlarger{\sum}_{r_n\in\mathlarger{\mathtt{R}^{s_n}_n}} P(r_{n+1}|r_n,s_n)\; P(r_n,s_n).
\end{align}
In each generation, the random variables $r_n$ and $s_n$ are independent, then $P(r_n,s_n)=P(r_n)P(s_n)$, and from \eqref{1} we have:
\begin{align}\label{11}
P(r_{n+1})\;=\underset{r_n\in\mathlarger{\,\mathtt{R}_n}}{\mathlarger{\sum}}\; T(r_{n+1}|r_n)\; P(r_n),
\end{align}
where 
\begin{align}\label{111}
T(r_{n+1}|r_{n}):=\mathlarger{\sum_{s_n=0,1}}\; P(r_{n+1},s_n|r_n)  \;\mathbf{1}_{s_n}(r_n),
\end{align}
and also $\mathbf{1}_{s_n}(r_n)$ is an indicator function, defined as $\mathbf{1}_{s_n}(r_n)=1$, if $r_n\in\pmb{\mathtt{R}}_n^{s_n}$ and $\mathbf{1}_{s_n}(r_n)=0$ otherwise. The expression \eqref{11} shows the markovian property of this stochastic  process. 
The transition elements $T(r_{n+1}|r_n)$ and initial condition govern the time evolution of the probability distribution $P(r_n)$. In Figure \eqref{diag2} the initial arrows connecting the nodes $s_i$ and $r_i$ with $r_{i+1}$ represent the possibility to \textit{throw} a random number $s_i$, in $i-$generation, in order to choose the sample space for the next random number $r_{i+1}$, given the number $r_i$. The same sentence is also valid for the arrows connecting the nodes $s_n$ and $r_n$ with $r_{n+1}$. By definition of biparental reproductive species, we have $r_0=2$. In this work we also assigned $r_1=2\,r_0$ directly, without any second random number $s_0$ in generation $n=0$. In other words, the initial condition for the simulation is given in $i=1$: $r_1=4$, the \textit{grandparents} generation.

We described the general algorithm with \textit{initial condition} at $i-$generation, $i=0,1$, denoting by $r_i=2^{i+1}$, trivially the expected value at this generation is $E(r_i)=2^{i+1}$.

We can obtain a recurrence equation for \textit{growth} and \textit{soft decay} regimes, presented in Section \ref{parameters-sec}. In that cases the algorithm establishes that 
\begin{equation}
P(r_{n +1}|\,r_n,s_n=0)=\delta_{r_{n+1},2\,r_n}
\end{equation}

Using this conditions on all the terms that $s_n=0$ of expression \eqref{1}, to reduce the expression  \eqref{exclusive expected} lead us to 
\begin{align}\label{3}
E(r_{n+1})&=2\,q_nE(r_n)+\xi_{n+1}.
\end{align}
This is a \textit{non-homogeneous linear recurrence equation}, valid from $n\geq i$, where $\xi_{n+1}$ is the in-homogeneity term given by
\begin{equation}\label{Inhomogen}
\xi_{n+1}:=\mathlarger{\sum}_{r_{n},r_{n+1}}
r_{n+1} P(r_{n+1}|\;r_n,s_n=1)\, P(r_n)\,p_n,
\end{equation}
where $r_n$ and $r_{n+1}$ runs over $\mathtt{R}_{n}$ and  $\mathtt{R}_{n+1}$, respectively.

This equation contains the trivial case: $p_n=0$ (i.e. $q_n=1$) for all $n\geq i$, corresponding to \textit{no inbreeding} condition in all generations. In this situation the \textit{non-endogamic} solution of \eqref{3} is 
\begin{equation}\label{NonEnd}
E(r_{n})=2^{n+1}.
\end{equation}

We can express the general solution of \eqref{3}, by iteration 
until $i-$generation, as follows
\begin{align} \label{InHomogSolution}
E(r_n)&=2^{n-i}\,q_{n-1}\cdots\, q_{i} E(r_i)+ \nonumber\\
&\\
&+\mathlarger{\sum}_{k=i+1}^n\;2^{n-k}\frac{q_{i}\cdots\, q_{n-1}}{q_{i}\cdots\, q_{k-1}}\;\xi_k,\nonumber
\end{align}
where $n\geq i+1$. Again if we take $q_k=1$ (e.g. $\xi_k=0$), for all $k$, in \eqref{InHomogSolution} we recover the \textit{non-endogamic} (or trivial) solution \eqref{NonEnd}.

The \textit{first term} of \eqref{InHomogSolution} corresponds to the case in which we have no inbreeding in all previous generations to $n$ (until $i-$generation), because the product $q_i\cdots\, q_{n-1}$ is just equal to this intersection probability
\begin{equation*}
P(s_i=0,\,\cdots,s_{n-1}=0)=P(s_i=0)\cdots P(s_{n-1}=0).
\end{equation*}
In the \textit{second terms} of \eqref{InHomogSolution} we have the quotients of $q_k's$ that are just equal to this intersection probability
\begin{equation*}
\frac{q_{i}\cdots q_{n-1}}{q_{i}\cdots q_{k-1}}=P(s_k=0,...,,s_{n-1}=0),
\end{equation*}
for $k=1,\,\cdots, n-1$ and is equal to 1, for $k=n$. 

The simulation used in this work is based on the case that $q_n=\frac{1}{2}$, for all $n\geq i$.
\begin{equation} \label{InHomogSolutionPaper}
E(r_n)=E(r_i)+\mathlarger{\sum}_{k=i+1}^n\;\xi_k.
\end{equation}
where the inhomogeneous terms $\xi_k$ can be written as 
$\xi_{k}=E(r_k|s_{k-1}=1)p_n$ and also we can simplify even more  $\xi_k=E(r_k,s_{k-1}=1)$.

Continuing with the last regime, for the \textit{hard decay} defined on Section  \ref{parameters-sec}, the algorithm establishes that $r_{n+1}$ is a uniformly distributed in $[2,\alpha_N]$, from \eqref{exclusive expected} we simply have: 
\begin{equation}\label{exclusive expected2}
E(r_{n+1})=q_n \frac{\alpha_N+2}{2} + \xi_{n+1}.
\end{equation}
because $E(r_{n+1}|s_n=0)=(\alpha_N+2)/2$ at this regime, i.e. is the first raw moment of random variable uniformly distributed in $[2,\alpha_N]$.

On the other hand, we can give a more satisfactory description in terms of high order moments of  $P(r_n)$. We obtained an expression for \textit{growth} and \textit{soft decay} regimes, from the moment generating function associated to $P(r_n)$.  We proved that
\begin{align}\label{higher moment recurrence}
E(r_{n+1}^{k})=2^k q_n\,E(r_{n}^{k})+\zeta_{n+1}^k
\end{align}
where $\zeta_{n+1}^k:=E(r_{n+1}^k,s_n=1)$. The recurrent equation \eqref{higher moment recurrence} has the structure such as the equation \eqref{3} and also shares the same kind of solution of equation \eqref{InHomogSolution}.

For the \textit{hard decay} regime and \eqref{exclusive expected} we obtained
\begin{align}\label{higher moment recurrence2}
E(r_{n+1}^{k})= \frac{q_n}{k+1}\sum_{l=0}^{k}2^{k-l}\alpha^{l}_N+\zeta_{n+1}^k
\end{align}
where $\zeta_{n+1}^k:=E(r_{n+1}^k,s_n=1)$ and $E(r^k_{n+1}|s_n=0)$ is the $k-$order raw moment of random variable uniformly distributed in $[2,\alpha_N]$.

We can even translate the center of the moment of $k-$order, assuming that we know all the previous moments  $\{E(x^l):\,l=0,1,\cdots\,k \}$, using $(x-c)^k=\sum_{l=0}^k {k \choose l}(-1)^{k-l}c^{k-l} x^l$, we have  
\begin{equation}
E[(x-c)^k]=\sum_{l=0}^k {k \choose l}(-1)^{k-l}c^{k-l} E(x^l),
\end{equation}
This expression can be useful to express $E[(x-E(x))^k]$ as a linear combination of power of $E^n(x)$ and $E(x^n)$ where $n=0,1,\cdots\, k$, in order to obtain the $k-$order moment centered around the mean value $E(x)$. 

This theoretical approach allow us to continue with the refinement of free parameters that the algorithm has, observing the analytical behavior of the solution, and in more general terms, the structure of the evolution equation. Regarding this, we want to add that there are also other ways to conduce the convergence of the tree. 

Other alternative is to use only a negative exponential distribution ($\sim \lambda_n e^{-\lambda_n r_n}$) for the whole \textit{decay interval} $[N_\mu,N]$ and use its parameter $\lambda_n$ to control the endogamy degree. This alternative exempts us from considering a cut generation, $N_\gamma$, a priori; there is no need to subdivide the \textit{decay interval} into \textit{soft} and \textit{hard}.

This mathematical approach can include the case where the random variables $\{r_n\}$ are continuous. This can be useful in order to simplify the calculations, since sometimes it turns out that sums are more complicated to treat than integrating.

In the following section we performed a comparison between the results of the proposed algorithm and the results of the theoretical paper \cite{nuestro-paper}.

\section{A comparison with the theoretical toy model}\label{compare}


We used the mean value of the number of ancestors presented in \cite{nuestro-paper} to compare it against data obtained with the computational model. The expression for the mean value of the number of ancestors at generation $n$, namely $\alpha(n)$, obtain in \cite{nuestro-paper} is given by:
\begin{equation}
\alpha(n)=2^{n+1}-\beta(n),
\end{equation}
\label{equ:ancestros}
where $\beta(n)$ represents the mean value of individuals who are outside to the set of ancestors, with respect to the maximum possible number of ancestors in each generation $n$, in this case is equal to $2^{n+1}$. Also 
\begin{equation}
\beta(n)=2^{\mathfrak{a}n+\mathfrak{b}}\langle X(n)\rangle
\end{equation}
is the product of expected value associated to the \textit{diluted process}, $\langle X(n)\rangle$, modulated by $2^{\mathfrak{a} n+\mathfrak{b}}$, \cite{nuestro-paper}.  Explicitly $\langle X(n)\rangle$ is equal to
\begin{equation}
\langle X(n)\rangle= e^{-2 n}[2 n \:I_1(2 n)+(2 n+\tfrac{1}{2}) I_0(2 n)]-\tfrac{1}{2},
\end{equation}
where $I_n(x)$ are the modified Bessel function \cite{Abramowitz}. 

The number of ancestors, $\alpha(n)$, depends on two parameters $\mathfrak{a}$ and $\mathfrak{b}$. 
If the expected value satisfies $\alpha(t_1)=\alpha_1$ and $\alpha(t_2)=\alpha_2$, for two generations $t_1$ and $t_2$ such that $t_1\neq 0\neq t_2$, the parameters $\mathfrak{a}$ and $\mathfrak{b}$ can be obtained by
\begin{align}\label{equ:5}
\mathfrak{a}&=\frac{1}{t_2-t_1}log_2 \left[\frac{2^{t_2+1}-\alpha_2}{2^{t_1+1}-\alpha_1}\;\frac{ \langle X(t_1)\rangle}{\langle X(t_2)\rangle}\right]\nonumber\\
\vspace*{1cm}&\\
\mathfrak{b}&=\frac{1}{t_2-t_1}\left\lbrace t_2\,log_2 \left[\frac{2^{t_1+1}-\alpha_1}{ \langle X(t_1)\rangle}\right]-t_1\,log_2 \left[\frac{2^{t_2+1}-\alpha_2}{ \langle X(t_2)\rangle}\right]\right\rbrace\nonumber
\end{align}
where $\alpha_i \leq2^{t_i+1}$, for $i=1,2$, to ensure a good definition of $\mathfrak{a}$ and $\mathfrak{b}$. These parameters can be related with the maximum number of ancestors in a given generation and the horizontal range \cite{nuestro-paper}.

\begin{figure}[!htbp]
\begin{center}
\includegraphics[totalheight=6.5cm]{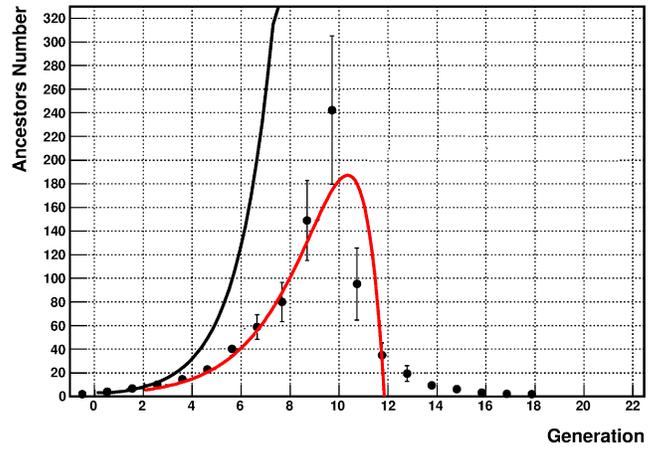}
\includegraphics[totalheight=6.5cm]{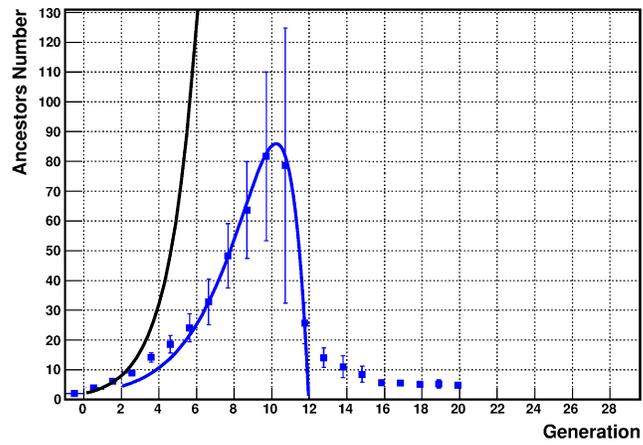}
\vspace{0.1cm}
\caption{(Color online). Mean number of ancestors for each generation. Average of 50 trees with the corresponding fit. The fit is valid up to 12 generations and the process ends a few generations late. TOP: uniform distribution. BOTTOM: negative exponential distribution.}
\label{fig:06}
\end{center}
\end{figure}

We used the equation for the mean number of ancestors depending on these parameters and performed a fit of the expression to our simulation data, in Figure \ref{fig:06}.

It is possible to obtain $\mathfrak{a}$ and $\mathfrak{b}$ parameters for any case that we want to study. The values for the fit parameter will make sense when experimental data would be use.

The Table I summarizes the values for $\mathfrak{a}$ and $\mathfrak{b}$ obtained for different distributions and the same final generations to end the process.

\begin{table}[ht!]
\centering 
\vspace{0.1cm}
\setlength{\arrayrulewidth}{0.25mm}
\bigskip
\begin{tabular}{|c|c|c|} 
\hline 
\multicolumn{3}{|c|}{$\quad$ \pmb{Uniform distribution} $\quad$}\\
\hline 
{$\quad\mathfrak{a}\quad$} & $0.997$  & $\pm 0.001$\\
\hline
{$\quad\mathfrak{b}\quad$} & $-0.73$   & $\pm 0.01$\\
\hline 
\multicolumn{3}{|c|}{\;\pmb{Negative Exponential distribution}\;}\\
\hline     
 {$\quad\mathfrak{a}\quad$ }& $0.956$  & $\pm 0.004$\\
\hline
 {$\quad\mathfrak{b} \quad$} & $\quad-0.25\quad$   & $\pm 0.4$\\
\hline
\end{tabular}
\bigskip
\vspace{0.1cm}
\caption{\small{Values of the fit parameters for two different cases corresponding to Figure 5 for trees of maximum generation $N=20$.}} 
\label{table:4}
\end{table}

The mean value of the number of ancestors in the set behaves with the generation in a similar way to the mean value obtained the theoretical model. This simulation could include more specific information of particular species or field studies with animals. Further it could be possible to combine the simulation with genetic algorithms to obtain a powerful tool to trace the combination of genes through the history of a particular specie. 

Even when we can obtain close results fitting the theoretical model parameters, our simulation is an improvement on the first theoretical model, in the sense that it is possible to modify and chose a certain distribution to model animal preferences in mate selection. Also the case studied in \cite{nuestro-paper}, that is a theoretical model in particular conditions, could not reproduce a slow decreasing of ancestors number as it is expected in a soft transition in consecutive generations. As we showed in \cite{nuestro-paper} the result of this model can be improved with a different choice of the gauge function $\lambda(t)$.	
In this algorithmic model, we can simulate an specific behavior in animal mate preference, according to the branch probability $q_n$ and the transition probabilities $P(r_{n+1}|r_n,s_n)$. This algorithm propose a more robust model that the first one proposed in \cite{nuestro-paper}. 

\section{Conclusions and further work}\label{pepe}

In present work we have built a simulation that is ruled by a small number of parameters and generates trees of ancestors. This simulation is based on a recursive algorithm and a random number generator. The model presented allows us to include animal mate preferences and to build more realistic trees than the previous model presented in \cite{nuestro-paper}.

This more sophisticated model allows us to include biological considerations represented through the distribution and its parameters. 

Different reproductive behavior means consider for instance one male mating with several females in a group or sibling selection. These behavior examples could be changed via the percent rate in this second generator $s_n$ from Section \ref{sofisti}. 

Even when this model has many empirical elements, it is a more realistic simulation of the trees that our previous version. In this regard, the simulation presents an opportunity to explore and discuss the elements of the mathematical description in a process of stochastic nature, in this case with markovian properties.

To develop the simulation, we used the available tools learned from the field of physics to generate a flexible dynamical model that could be used for biologist to compare and make predictions with real animal data.

The model mainly uses the hypothesis of a certain degree of inbreeding as the key to the development of any animal population of sexual reproduction. We leave open the question of which is the degree of endogamy required to a healthy population development but we claim that no population could develop without an certain level of endogamy within.

Our model could be used as a powerful tool in order to contribute in ecology and biology studies by using empirical data collected from an animal behavior of a any population to constraint the parameters of the model and make predictions. Additionally, the simulation as well as the algorithm can be used to describe other biological or physical systems with similar dynamics. Such models has been used before in that task \cite{18-nuestro}.

Another interesting point is regarding how topology could affects the evolution of a population \cite{19-nuestro}.

Some open questions related to the nature of the process have raised from the developed simulation, for instance: How many generations makes sense to follow a tree of ancestors? How to chose where to stop the tree? What happens with the human case? Could be possible with this model shows differences between animal groups of sibling
species.

We have developed a model through an algorithm that allows to gain understanding of the future experiment. We showed how relatively simple selections of this two distributions: the one-dimensional $P(s_n)$ and the conditional $P(r_{n+1}|s_n,r_n)$ allow us to describe the beginnings of a phenomenology of the concrete process involved in the ancestors trees formation. Such phenomenology can be enriched as long as these probability distributions become more complex.

Further work will include a development of an algorithm version in python code, with parallel programing improvements.

\vskip6pt
\section*{ACKNOWLEDGMENTS}

This work was supported by CONICET and UNQ institutions. We thank the scientists from the biology field who encourage us to keep working in biological models. We thank also our colleagues Graciela Molina and Pablo Alcain for their contribution to this work. Special mention to Cristina J. for his critical objections and text style corrections. Finally, we want to mention to Micaela Moretton, Mar\'ia Clara Caruso, and Gabriel Lio for always give us personal support. During the development of this work we have a member to include in the Caruso family tree (little Lucia) and we want to dedicate present work to her.






\end{document}